\documentclass[a4paper,12pt,reqno,superscriptaddress]{revtex4}

\usepackage[centertags]{amsmath}
\usepackage{amsfonts}
\usepackage{amssymb}
\usepackage{amsthm}
\usepackage{newlfont}
\usepackage{stmaryrd}
\usepackage{mathrsfs}
\usepackage{euscript}
\usepackage{graphicx}
\usepackage{enumerate}
\usepackage{tikz}
\usepackage{pgf}
\usetikzlibrary{positioning,fit,calc}
\usetikzlibrary{arrows,automata}
\usepackage{wrapfig}
\usepackage{subfigure}
\usepackage{amscd}
\usepackage{hyperref}

\theoremstyle{plain}

\theoremstyle{definition}

\theoremstyle{remark}

\newcommand{\bbR}{{\mathbb R}}

\newcommand{\bbZ}{{\mathbb Z}}

\newcommand{\opunit}{\text{1}\kern-0.22em\text{l}}

\DeclareMathAlphabet{\mathpzc}{OT1}{pzc}{m}{it}

\newcommand{\id}{\textrm{d}}

\begin{document}

\title{{\bf On the kinetics that moves Myosin V\\ }}

\author{Christian Maes and  Winny O'Kelly de Galway\\
Instituut voor Theoretische Fysica, KU Leuven}

\begin{abstract}
Molecular motor proteins such as Myosin V, Dynein or Kinesin are no ratchets, at least not with a flashing asymmetric potential; the crucial asymmetry is in the dynamical activity. We make that explicit in terms of a simple Markov model, emphasizing the kinetic (and non-thermodynamic) aspects of stochastic transport.  The analysis shows the presence of a fluctuation symmetry in that part of the dynamical activity which is antisymmetric under reversal of trailing and leading head of the motor.  The direction of the motor motion is determined by it.
\end{abstract}
\maketitle
\baselineskip=20pt
\section{Introduction}
In Disney's version of Rudyard Kipling's Jungle Book, Mowgli attacks Shere Khan and ties a flaming branch to the tiger's tail. Khan frenetically flees in panic from the fire.  Imagining the tiger king standing on a one-dimensional track, being able only to move forward or backward, there are good reasons for the tiger to typically move forward.  After all, its tail is on fire and there is an asymmetry in the tiger configuration, having the tail on its backside.  Moreover, the tiger is alive --- dead tigers do not move forward even with a burning tail.  Those then, we claim, are analogies for what makes Myosin V typically move forward on actin tracks by the consumption of adenosine triphosphate (ATP).\\

Myosin V is a two-headed motor moving in a head-over-head fashion. Its motion has been largely explored experimentally, in particular thanks to its long (24 nm) lever arms and big displacement (36 nm per step), \cite{crapna,vale}. Rather detailed mechanochemical aspects of the motor have been verified and visualized, \cite{crapna,craig}. The cycles of subsequent detachment and attachment with hydrolysis of ATP allow here some simplification and make it possible to concentrate the discussion on essential aspects of molecular motor motion. In the present paper we do not propose new chemical or structural aspects to the main cycle of Myosin V motion except for focusing on the question {\it why} it is selected from the various thermodynamic possibilities. For major inspiration we refer here in particular to the papers by Dean Astumian, \cite{astu,rdas} where the kinetics is discussed with emphasis on the microscopic reversibility.  We will formalize some of those ideas here using simple Markov models and, while fully agreeing with the main points, we want to emphasize instead the frenetic aspects in the motor's working \cite{fren}. It also leads to the prediction of a fluctuation symmetry in these kind of motors which does not involve the entropy flux (in contrast with the standard fluctuation theorem for Markov jump processes \cite{lebsp,fluc}) but involves the dynamical activity, \cite{maar,woj}.  In that way Myosin V offers a nice illustration of the role of time-symmetric aspects and how they can play a role in constructing nonequilibrium statistical mechanics.\\
Before we start however with the more detailed analysis of the stepping of Myosin V, we present in the next section more elementary Markov toy models to illustrate key points that will be applied further on.\\
We first relate the Myosin V stepping with a Markov model in Section \ref{myo}.  We then have to present the broader theoretical framework of dynamical ensembles, in Section \ref{den}, to arrive at our main findings for Myosin V in Section \ref{main}.

\section{Preliminary abstraction}

To start with the simplest nonequilibrium model, suppose a three-state Markov process with transition rates 
\begin{eqnarray*}
k(1,2) = ae^{q},\quad k(2,3) = be^q,\quad k(3,1) = ce^q,\\
k(1,3) = ce^{-q},\quad k(3,2) = be^{-q},\quad k(2,1)=ae^{-q} 
\end{eqnarray*}
parameterized by external field $q\geq 0$ and numbers $a>b>c> 0$. If $q=0$, there is detailed balance with the equilibrium distribution being uniform on $\{1,2,3\}$, whatever the $a,b,c$.  From the moment $q>0$, the asymptotic behavior is that of steady nonequilibrium where the driving $q$ does not as such distinguish between the three states.  However, the prefactors $a,b,c$, while time-symmetric over the jumps $1\leftrightarrow 2\leftrightarrow 3$,  now determine the stationary condition.  As is easily seen, for large $q$ the stationary distribution concentrates on state 3.  The reason is that the escape rate from state 3 is the least; the system will spend most of its time in 3; see \cite{JPA,heatb} for a more precise and general version of the statement.  That concerns a static property, but there is a dynamical analogue, which is next.\\

  Suppose next a more general Markov jump process with states $x,y,\ldots \in K$ modeling an open physical system in contact with one or more sufficiently separated thermal equilibrium reservoirs.  For all transitions $x\rightarrow y$ we assume known  the corresponding calorimetric entropy flux $S(x,y)=-S(y,x)$ into the environment.  Physical modeling (the microscopic reversibility as emphasized in \cite{rdas}) then requires that the Markov process satisfies local detailed balance in the sense that
 $k(x,y) >0$ not only implies $k(y,x)>0$ but also that
\begin{equation}\label{ldb}
\frac{k(x,y)}{k(y,x)} = \exp S(x,y)/k_B
\end{equation}
for all  transition rates \cite{tas,mn}. (We come back to that condition in Section \ref{den}.)  It allows to compute the entropy flux over a path or trajectory in state space by multiplying the factors in \eqref{ldb} according to the jumps that have taken place.\\
Suppose then that for a certain state $D$ there are cycles $\omega^{(1)}: D\rightarrow x_2^{(1)}\rightarrow \ldots x_n^{(1)}\rightarrow D$ and $\omega^{(2)} : D\rightarrow x_2^{(2)} \rightarrow \ldots x_m^{(2)} \rightarrow D$ on $K$ having the same entropy flux,
\begin{equation}\label{oom1}
S(\omega^{(1)}) =  S(D,x_2^{(1)}) +\ldots S(x_n^{(1)},D)\, =  \,S(D,x_2^{(2)}) +\ldots S(x_m^{(2)},D)  =S(\omega^{(2)})
\end{equation}
(Residence or dwelling times in each state of the cycle do not matter for the entropy flux.)  Similar to our previous example, \eqref{oom1} implies that the dissipative driving aspect does not distinguish between the two cycles; whence we ask which one will be taken more often?
What decides whether $\omega^{(2)}$ or rather $\omega^{(1)}$ will be preferred by the system?  The answer is a version of Landauer's principle {\it motion out of noise}, \cite{land,land2,heatb} to be recalled in Section \ref{den}, which is, so we claim, the heuristics behind much of the kinetics of at least some class of molecular motors. The following sections will make that precise. Let us however already make that more specific by realizing the above  on six states $K=\{D,x,v,T,w,y\}$ in precisely the set-up that we will need for Myosin V.

\begin{figure}[h!]
\centering
 \begin{tikzpicture}
[-,>=stealth',shorten >=1pt,auto,node distance=3cm,
   thick,main node/.style={circle,fill=blue!10,draw,font=\sffamily\small\bfseries},main node/.style={circle,fill=blue!10,draw,font=\sffamily\small\bfseries},state/.style={font=\sffamily\small\bfseries}]

  \node[state] (1) {v};
  \node[main node] (2) [below left=0.8 cm and 2.2 cm of 1] {D};
  \node[state] (3) [below right=0.8 cm and 2.2 cm of 2] {w};
  \node[main node] (4) [below right=0.8 cm and 2.2 cm of 1] {T};
  \node[state] (5) [above= 0.8 cm of 1] {x};
  \node[state] (6) [below =0.8 cm of 3] {y};
    \node[] (7) [ left of=2] {};
     \node[] (8) [right of=4] {};
    \path [dotted] (7) edge  (8);
  \path[every node/.style={font=\sffamily\small}]
     (2) edge  [green,->] node {} (1)
    (1)   edge [green,->] node {} (4)
     (4)   edge [green,-> ]node {} (6)
      (6)  edge [green,-> ]node {} (2)
  (3) edge [red,->](4)
  (2) edge[red,->] (3)
  (5) edge [red,->](2)
  (4) edge [red,->] (5);
\end{tikzpicture}
\caption{Abstraction of Myosin V as a six state Markov process with mirroring cycles over $R_1$ (green and counter-clockwise) and $R_2$ (red and clockwise).}
\label{fig1}
\end{figure}
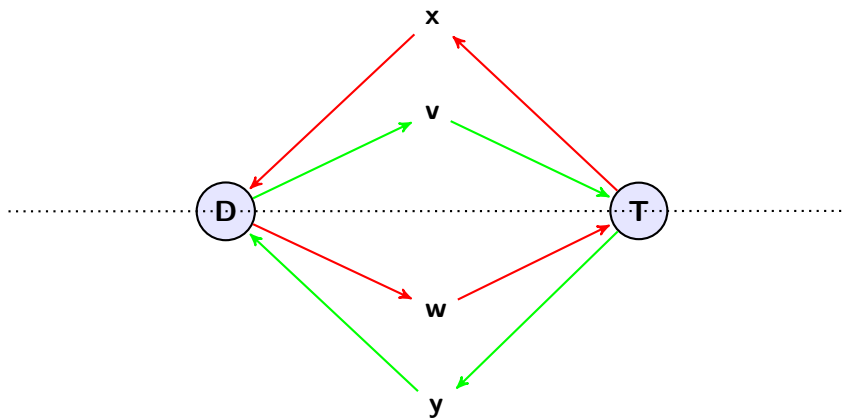

Here are the rates, first for the transitions $D\rightarrow v \rightarrow T\rightarrow y\rightarrow D$,
\[
k(D,v) = a,\quad k(v,T) = \psi_1,\quad k(T,y) = \psi_2\,e^{s_2},\quad k(y,D) = d\,e^{s_3}
\]
and with the rates for the inverse transitions decided by local detailed balance:
\[
k(v,D) = a\,e^{s_0},\quad k(T,v) = \psi_1 \,e^{-s_1},\quad k(y,T) = \psi_2,\quad k(D,y) = d
\]
The $s_0, s_1, s_2, s_3$  are known entropy fluxes  (from now on always per $k_B$), so we suppose, over the corresponding jumps, with choices $s(v,D)=s(w,D)=s_0,\, s(v,T)=s(w,T)=s_1,\, s(T,y)=s(T,x)=s_2,\, s(y,D)=s(x,D)=s_3$.
Similarly, for transitions $D\rightarrow w \rightarrow T\rightarrow x\rightarrow D$,
\[
k(D,w) = b,\quad  k(w,T) = \psi_1,\quad k(T,x) = \psi_2 \,e^{s_2}\quad k(x,D) = c\,e^{s_3} 
\]
and, for their reverse,
\[
k(w,D) = b\,e^{s_0},\quad  k(T,w)= \psi_1\,e^{-s_1},\quad k(x,T) = \psi_2,\quad k(D,x) = c
\]
The numbers $a,b,c,d,\psi_1,\psi_2$ are prefactors that we will call reactivities in what follows.  They reflect kinetic aspects of the transitions.\\
We consider the two rings in Fig.~\ref{fig1} which touch each other at the states $D$ and $T$.
Let us call $R_1=(D,v,T,y,D)$ and $R_2= (D,w,T,x,D)$; they are each other's reflection over the dotted horizontal line in Fig.~\ref{fig1}. We are interested in knowing whether the system will prefer cycling over $R_1$ or over $R_2$ (the first being clockwise and the latter having the counter-clockwise direction).  To make it interesting we have chosen the rates above so that the entropy fluxes $S_1$ and $S_2$ respectively for going through $R_1$ and for going through $R_2$, are exactly equal:
\begin{equation}\label{0123}
S_1 = -s_0 + s_1 + s_2 +s_3 = S_2  
\end{equation}
 and similarly, for the reversed cycles in each ring.  The equality \eqref{0123} follows from applying \eqref{ldb}, e.g.
 \[
S_1 = \log \frac{k(D,v)\,k(v,T)\,k(T,y)\,k(y,D)}{k(D,y)\,k(y,T)\,k(T,v)\,k(v,D)} 
 \]
Eq. \eqref{0123} in fact realizes the equality \eqref{oom1} with $n=m=4$ for $\omega^{(1)}$ running clockwise over $R_1$ and $\omega^{(2)}$ running counter-clockwise over $R_2$: $S(\omega^{(1)}) = S_1 = S_2 = S(\omega^{(2)})$.
It means in particular that the sign of the entropy flux does not determine here the direction of motion: if $S_1>0$ so is $S_2>0$, and {\it vice versa}, but they correspond to cycling in opposite directions.  From the perspective of dissipation the system will prefer orbiting with positive entropy flux but that can be equally realized by clockwise (via $\omega^{(1)}$) or by counter-clockwise (via $\omega^{(2)}$) motion.  We conclude that what truly decides the direction of motion here is not-entropic; it is frenetic, \cite{fren,land2}.\\
  The asymmetry between the two cycles indeed resides in the difference in reactivities, the symmetric prefactors in the transition rates, e.g. $a\neq b$ for the transition rates $D\rightarrow v$ {\it versus} $D\rightarrow w$, but more specifically the system will prefer ``the least noisy trajectory'' when giving the choice between two that have the same entropy flux.  That can be quantified using the notion of dynamical activity, as will follow in Section \ref{den}. Note however before going any further that in the case of (global) detailed balance, when $s_0=s_1+s_2+s_3$ (as necessary and sufficient condition), then the discrepancy $a\gg b$ or $c\neq d$ etc.  plays no role and the stationary regime would be equilibrium-dead showing no systematic orbiting at all.
We recognize the three features ((1) asymmetry in the (2) dynamical activity (3) without detailed balance) associated to Shere Khan featuring in the beginning of the Introduction.  While the above examples remain rather abstract, they are simple and already collect the most important nonequilibrium statistical mechanical aspects of Myosin V motion.

\section{Myosin V stepping}\label{myo}

A  mammalian myosin V protein moving along an actin track hops stochastically in steps equal to about 72 nm along an actin track, \cite{rief}.
 One typically sees via electron microscopy or using single molecule microscopy that the lever arm of the trailing head rotates between the rearward- and forward-pointing positions, \cite{vale}. The trailing head appears then to be pushed past the stationary forward head after which it settles to bind on  a new actin site.
  Roughly one step corresponds to the consumption of one ATP molecule; yet this could also be a step backwards and there is deviating behavior. That already suggests that the direction in which the motor is moving is not determined by the entropy fluxes; i.e., the second law perhaps influences but does not decide here the direction of transport.  Nevertheless, whatever mechanism or complicated combination of chemistry, directed motion albeit diffusion is not possible in steady equilibrium, so some nonequilibrium aspect must be coming in. In order to understand the situation better we must look at some pieces of phenomenology.\\

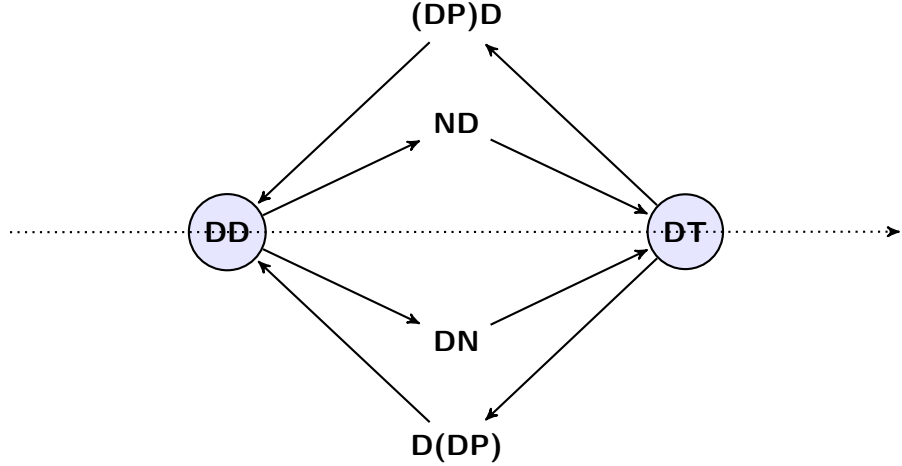
\begin{figure}[h!]
\centering
\begin{tikzpicture}[->,>=stealth',shorten >=1pt,auto,node distance=3cm,
  thick,main node/.style={circle,fill=blue!10,draw,font=\sffamily\small\bfseries},state/.style={font=\sffamily\small\bfseries}][]

  \node[state] (1) {ND};
  \node[main node] (2) [below left=0.8 cm and 2.2 cm of 1] {DD};
  \node[state] (3) [below right=0.8 cm and 2.2 cm of 2] {DN};
  \node[main node] (4) [below right=0.8 cm and 2.2 cm of 1] {DT};
  \node[state] (5) [above= 0.8 cm of 1] {(DP)D};
  \node[state] (6) [below =0.8 cm of 3] {D(DP)};
    \node[] (7) [ left of=2] {};
     \node[] (8) [right of=4] {};
    \path [dotted] (7) edge  (8);
  \path[every node/.style={font=\sffamily\small}]
     (2) edge  [->] node {} (1)
    (1)   edge [->] node {} (4)
     (4)   edge [-> ]node {} (6)
      (6)  edge [-> ]node {} (2)
  (3) edge [->](4)
  (2) edge[->] (3)
  (5) edge [->](2)
  (4) edge [->] (5);
\end{tikzpicture}\caption{The transition scheme of Myosin V, realizing the Markov process of Fig.\ref{fig1}.}
\label{fig:M1}
\end{figure}

The stepping of Myosin V is written as a completely coupled cycle; see Fig.~\ref{fig:M1}. Either ADP (denoted by D) or ATP (denoted by T) is bound to the heads of the protein. We can imagine that the motor has basically two states between which it switches: either both front and rear head are bound with ADP (state DD) or one is bound to ATP (state DT). In order to switch between those two states there are four different channels of which however only two are thermodynamically different.\\
 We start from state DD and  ADP can be released from the motor leaving a non-occupied site, denoted by (N). We are then at state ND or at state DN in Fig.\ref{fig:M1}.  At this point ATP comes in and binds to the non-occupied site detaching it from the actin track, entering state DT. Since the thermodynamics of releasing ADP  from the rear or front head ($s(DD,DN) = s(DD,ND)=:-s_0$), and similarly of ATP binding to the rear or to the front head are equal ($ s(DN,DT) = s(ND,DT)=: s_1$), there are two paths from $DD$ to $DT$ along which the released heat and hence the entropy flux are identical: $s(DD,DN) + s(DN,DT) = s(DD,ND) + s(ND,DT) = s_1-s_0$.\\
A second thermodynamic-type of channels involves synthesizing or hydrolizing ATP into ADP and inorganic phosphate Pi. From the ADP and Pi as an intermediary state, be it $(DP)D$ or $D(DP)$ in Fig.\ref{fig:M1}, Myosin loses its phosphate to reattach to the actin track, and we are back in state DD.  The entropy fluxes are
  $s(DT,(DP)D) = s(DT,D(DP))=: s_2$ and $s((DP)D,DD) = s(D(DP),DD)=: s_3$.\\
The entropy fluxes can all be identified.  We have reservoirs in which inorganic phosphor Pi, ADP and ATP molecules are free and influence the rates of attachment and detachment via their respective chemical potentials $\mu_{Pi},\mu_D$ and $\mu_T$ in  the motor's environment (at uniform inverse temperature $\beta$).  That implies $s_0 =\beta \mu_D, s_1=\beta\mu_T, s_3=-\beta\mu_{Pi}$.  On the other hand for the hydrolysis/synthesis $ATP \leftrightarrow ADP + Pi$ there corresponds an entropy flux $\beta\Delta \mu$ which determines $s_2= \beta\Delta\mu >0$.     We have no driving (detailed balance) when $s_2= s_0-s_1-s_3$ or, $\Delta \mu = \mu_D  - \mu_T + \mu_{Pi}$ which is avoided by the abundance of ATP in the motor's environment.
The excess concentration of ATP implies that ATP binding is promoted, ATP dissociation is less likely than in equilibrium, and similarly Pi dissociation wins with respect to Pi association.  That is formalized in the inequality $\dot S := s_1+s_2+s_3-s_0 = \Delta \mu  -\mu_D  + \mu_T - \mu_{Pi} > 0$.

The reader surely sees the similarities between Fig.~\ref{fig:M1} and Fig.~\ref{fig1}; one can in fact estimate quite well the different stages in one typical forward step. At high ATP concentrations, the dwell time distribution of $D$ appears to be exponential, confirming a Markov approximation with an escape rate  of 12 transitions per second \cite{rief,forkey}.  Repeating the notation of the previous section for the four channels we have rates
\begin{eqnarray}\nonumber
k(DD\rightarrow ND \rightarrow DT)= a\, \psi_1\quad
k(DD\rightarrow DN \rightarrow DT)=b \,\psi_1 \\ \nonumber
k(DD\rightarrow (DP)D \rightarrow DT)= c\, \psi_2,\quad
k(DD\rightarrow D(DP) \rightarrow DT)= d\,\psi_2\label{4ch}
\end{eqnarray}

For our most important point about the phenomenology we stress that there is an important asymmetry between rear (trailing) and front (leading) heads.   Because of the conformation and the asymmetry of the actin filament, what is rear and what is front head can always be distinguished.  The transition $DD \leftrightarrow ND$ (dissociation or association of ADP at the rear head) is much faster then the transition $DD \leftrightarrow DN$ (dissociation or association of ADP at the front head). That is described in  Eq.4 in \cite{crapna} and is quite appropriately called there the
kinetic asymmetry.
 In other words, the chemical difference between rear and front head manifests itself in the easier release and binding of ADP at the rear head.  It can in fact be monitored experimentally that ADP release triggers the step, but almost without exception it is the trailing head that dissociates and moves during the step \cite{yildiz}.  The same is true but with front and rear exchanged, for the release and binding of Pi in the transitions $DD\leftrightarrow (DP)D$ {\it versus} $DD\leftrightarrow D(DP)$. Formally, and referring to the notation for the transitions in the six state Markov process in the previous section or to \eqref{4ch} for the rates in the abstract model Fig.~\ref{fig1},  $a\gg b$, $d\gg c$ for (rear {\it versus} front) differences in time-symmetric reactivities. That rear--front asymmetry is however purely time-symmetric and has nothing to do with the nonequilibrium condition although, as in Shere Khan's agony, it is the latter that makes the first relevant.\\

 Observe for the rest that the transitions $x,y\leftrightarrow T$ and $v,w\leftrightarrow T$ in Fig.~\ref{fig1} have respectively $k(x,T)=k(y,T)$ and $k(v,T)=k(w,T)$ (mirror symmetry), which means a front {\it versus} rear head symmetry in Fig.~\ref{fig:M1} making $k(ND,DT)=k(DN,DT)$ and $k((DP)D,DT)=k(D(DP),DT)$ (and similarly for the reversed rates).

\section{Dynamical ensembles}\label{den}
To continue we discuss a number of points that have partially been mentioned before and are important for the discussion of the working of the molecular motor.\\
The idea above was to model the Myosin V steps in terms of a six-state Markov process (or, what amounts to the same, a two-state Markov process with four channels) introduced in Fig.~\ref{fig1}.  Such a Markov process gives rise to a probability on trajectories or paths over time-windows $[0,t]$, the so called path-space distribution $P$.  (We will not write the dependence on the end-time $t$.)  Paths $\omega = (x_u, 0\leq u\leq t)$ have states $\omega_u\equiv x_u\in K$ at time $u$, constant during waiting or residence times $u_{i+1}-u_i$ (where the process remains in  state $x_{u_i}$) and they jump at times $u_i$ (where transitions $x_{u_{i-1}}\rightarrow x_{u_i}$ actually occur).
To describe that distribution $P$ it is useful to take a trivial reference process $P^0$ on $K$ where all the transition rates are just equal to one, and to define an action ${\cal A}$ on all allowed paths $\omega$ via
\[
P(\omega) = e^{-{\cal A}(\omega)}\,P^0(\omega)
\]
or, $e^{-{\cal A}(\omega)}$ is the density on path-space.  Clearly all the relevant information about the process now sits in the action ${\cal A}$ and should usefully translate the chemo-mechanical information in the transition rates.\\

To be more explicit about the action $\cal A$, we consider all the transitions $x\rightarrow y$ with $k(x,y)>0$ for the original process to which we give reference transition rates $k^0(x,y)=1$.  Clearly then, the corresponding process $P^0$ when randomly started will remain very random, and enjoying all possible temporal and spatial symmetries. For the true rates we use the parameterization
\begin{equation}
k(x,y)= \phi(x,y)\,\exp[\frac{1}{2}s(x,y)]\label{pa}
\end{equation}
that distinguishes the time-symmetric component $\phi(x,y)=\phi(y,x)>0$  from the antisymmetric $s(x,y) = -s(y,x)$.  That should also be done on the level of the action $\cal A = D- S/2$, to be decomposed in time-symmetric contribution $D$ and time-antisymmetric part $S$.\\

The condition of local detailed balance ensures that $S$ has a clear thermodynamic meaning: it makes the entropic component in the action $\cal A$,
\begin{equation}\label{so}
S(\omega) = \sum_{i} s(x_{u_{i^-}},x_{u_i})
\end{equation}
with the sum over jump times $u_i$ when $x_{u_{i^-}}\rightarrow x_{ui}$ as dictated by the path $\omega$  (we use $i^-\equiv i-1$), see e.g. \cite{tas,mn}.\\
Reactivities $\phi(x,y)$ are in principle given from reaction rate theory by a version of the Arrhenius--Kramers formula but can also be measured to some extent for Myosin V, \cite{craig}. 
Continuing with  the parameterization \eqref{pa} in the decomposition $\cal A = D- S/2$, we find what is called the frenetic component
\begin{equation}\label{dynsa}
D(\omega) = -\sum_{i} \log\phi(x_{u_{i^-}},x_{u_i}) + \int_0^t\id u \sum_{y\sim x_u} [k(x_u,y)-1]
\end{equation}
where the first sum is again over the jump times in the path over $[0,t]$ and the second sum is over all states $y$ for which $k(x_u,y)>0$.
The expression \eqref{dynsa} is an excess dynamical activity (also called the frenetic contribution \cite{fren,resp}) as it involves the time-integrated change in escape rates, together with the sum of reactivities $\log\phi(x,y)$ for all transitions $x\rightarrow y$ during $[0,t]$. 
We can also write \eqref{dynsa} as
\begin{equation}\label{dynsag}
D(\omega) = -\sum_{i} \log\phi(x_{u_{i^-}},x_{u_i}) + \sum_{x\in K}\xi(x)\, \tau_x^{[0,t]}
\end{equation}
That path-dependent dynamical activity plays an increasing role in studies of nonequilibrium statistical mechanics; see e.g. \cite{vW,chan,fren} for some examples.  Clearly, paths $\omega^{(1)}$ and $\omega^{(2)}$ for which $S(\omega^{(1)}) = S(\omega^{(2)})$ (equal entropy fluxes) and $P^0(\omega^{(1)}) = P^0(\omega^{(2)})$ (no preference according to the reference random process) will have different plausibility depending on the value of that frenetic term:
\begin{equation}\label{dw}
\frac{P(\omega^{(1)})}{P(\omega^{(2)})} = \exp [D(\omega^{(2)})- D(\omega^{(1)})]
\end{equation}
We will apply that in the next section.
It is a version of Landauer's principle {\it motion out of noisy states} \cite{land}, as the paths with lowest excess dynamical activity would be preferred.  More concretely the preferred path will visit states along which the reactivities $\phi(x,y)$ are larger and for which the time-integrated escape rates are smaller.\\

The decomposition above uses time-reversal $\Theta$ to split the action as $\cal A = D- S/2$ with $S\Theta=-S, D\Theta = D$.  Here
$\Theta \omega$ is the new path obtained from $\omega$ via $(\Theta\omega)_u=x_{t-u}$ for time $u\in [0,t]$. For our purposes it is interesting to use a second symmetry or symmetry transformation $\Gamma$ on paths $\omega$ which  commutes with time-reversal, $\Theta\Gamma= \Gamma\Theta$, and is also an involution, $\Theta^2= \Gamma^2= $ Id.  The symmetry $\Gamma$ will be a spatial symmetry in the application to Mysosin V, that exchanges the rear and the front heads.  In that spirit we also assume now that $S\Gamma = S$ (there is no thermodynamic difference between leading and trailing head), and we decompose further $D= F - H/2$ where
\[
F := \frac{1}{2}[D\Gamma + D],\quad  H := D\Gamma - D =  \cal A \Gamma - \cal A
\]
so that $H$ is antisymmetric under the transformation $\Gamma$ and symmetric under $\Theta$.  For paths 
 $\omega^{(1)}$ and $\omega^{(2)}$ for which $\omega^{(1)} = \Gamma\omega^{(2)}$ we obviously have now
\[
\frac{P(\omega^{(1)})}{P(\omega^{(2)})} = \exp H(\omega^{(1)})
\]
improving on \eqref{dw}.
Moreover, for all functions $f$,
\begin{eqnarray}\label{fsh}
\langle f(-H)\rangle &:=& \int {\cal D}[\omega] P^0(\omega) \exp\left[\frac 1{2}S(\omega) - F(\omega) + \frac 1{2}H(\omega)\right]\,f(-H(\omega))\nonumber\\
&=& \int {\cal D}[\omega] P^0(\omega) \exp\left[\frac 1{2}S(\omega) - F(\omega) - \frac 1{2}H(\omega)\right]\,f(H(\omega))\nonumber\\&=& \int {\cal D}[\omega] P^0(\omega) \exp\left[\frac 1{2}S(\omega) - F(\omega) + \frac 1{2}H(\omega)\right]\,e^{-H(\omega)}\,f(H(\omega))
\end{eqnarray}
which is a fluctuation symmetry $\langle f(-H) \rangle = \langle e^{-H}\,f(H)\rangle$.  As a consequence, by taking $f\equiv 1$, and by the convexity of the exponential function we get the positivity $\langle H\rangle \geq 0$ for the average value of $H$.  That follows the same reasoning as in e.g. \cite{fluc} for the fluctuation symmetry of the entropy flux $S$ with time-reversal $\Theta$ then replacing $\Gamma$.  For our two symmetries we thus have for all $h, \sigma \in \bbR$,
\[
\frac{\text{Prob}[H = h]}{\text{Prob}[H = -h]} = e^h, \quad \frac{\text{Prob}[S = \sigma]}{\text{Prob}[S = -\sigma]} = e^\sigma
\]
where the probabilities are with respect to the original Markov process that was given a hot (completely random)  start.  It also means that the paths where both $S$ and $H$ are positive will be preferred and that decides the direction of the molecular motor.

\section{Direction of the motor's motion}\label{main}

We go back to Myosin V and the cycle as depicted in Fig.~\ref{fig1} or Fig.~\ref{fig:M1}. For  a path $\omega^{(1)}$ that follows $R_1$ (the green cycle in Fig.~\ref{fig1}), then its time-reversal $\Theta\omega^{(1)}$ follows $R_1^c= (D,y,T,v,D)$.  Similarly, a path 
 $\omega^{(2)}$ that cycles over the red path $R_2$ in Fig.\ref{fig1}) has time-reversal $\Theta\omega^{(2)}$ cycling $R_2^c = (D,x,T,w,D)$.  In the notation of Section \ref{myo},
\begin{eqnarray}
R_1 &=& (DD,ND,DT,D(DP),DD),\quad R_1^c = (DD,D(DP),DT,ND,DD)\nonumber\\
R_2^c &=& (DD,(DP)D,DT,DN,DD), \quad R_2 = (DD,DN,DT,(DP)D,DD)\label{amon}
\end{eqnarray}

 The transformation $\Gamma$ reverses the front with the rear head, in the sense of reflection along the horizontal line in Figs.\ref{fig1} and \ref{fig:M1}: $(\Gamma\omega)_s= \widetilde{\omega_{s}}$ with $\tilde{x}=y,\tilde{v}=w$ so that for  $\omega^{(1)}: D\rightarrow v \rightarrow T \rightarrow y \rightarrow D$  we have $\Gamma\omega^{(1)} = \omega^{(2)}: D \rightarrow w \rightarrow T \rightarrow x \rightarrow D$.  To see which one is  preferred we  must find the sign of 
 \[
 H(\omega^{(1)}) = \log\frac{\phi(D,v)}{\phi(D,w)}\, I_{Dv} + \log\frac{\phi(y,D)}{\phi(x,D)} \,I_{yD}    + [\xi(w)-\xi(v)]\, \tau_v^{[0,t]}
 + [\xi(x)-\xi(y)]\, \tau_y^{[0,t]}
 \]
where $I_{Dv}$ and $I_{yD}$ is the number of transitions between states $D\leftrightarrow v$, respectively 
 $D\leftrightarrow y$.  In terms of the Myosin V stepping that means
 \begin{eqnarray}\label{ha}
 H(\omega^{(1)}) &=& \log\frac{a}{b}\,\, I_{DD,ND} + \log\frac{d}{c} \,\,I_{DD,D(DP)} \\&+ &  (b-a) e^{s_0}\, \tau_{ND}^{[0,t]}
  + (c-d)\,e^{s_3}\, \tau_{D(DP)}^{[0,t]}\nonumber
 \end{eqnarray} 
 which is also part of the crucial quantity that appears in Eq. 7 in \cite{astu}.
 Indeed, observations show that states $D(DP)$ and $ND$ are short lived  transition states and that  $a\gg b$, $d\gg c$, realizing thus the afore mentioned preference for dissociation or association of ADP at the rear head, and for the front head release and binding of Pi in the transitions $DD\leftrightarrow (DP)D$ {\it versus} $DD\leftrightarrow D(DP)$. That implies that  $H(\omega^{(1)})>0$, and hence it is preferred.\\
   To summarize, the fact that $a\gg b$, $d\gg c$ implies that the clockwise-path following $R_1$ instead of the counter clock-wise path following $R_2$ will be much preferred, or $\text{Prob}[R_1] > \text{Prob}[R_2]$.  In fact the path $\omega^{(1)}$  following $R_1$ is unique among the four types \eqref{amon} of cycles where both $H(\omega^{(1)})>0$ and $S(\omega^{(1)})>0$:  $H(R_1^c)= H(R_1) >0, H(R_2^c) = H(R_2) < 0,\,S(R_1^c)= S(R_2^c) < 0, S(R_1) = S(R_2) >0$,
from which we conclude that the cycle $R_1$ is the most plausible.  That means forward motion.  At the same time, the analysis of the previous section predicts a symmetry in the fluctuations of the quantity $H(\omega)$.

\section{Conclusions}
Myosin V  is a very well-studied molecular motor with basic biological functions such as retaining vesicles and organelles in the actin-rich periphery of cells.  It provides a perfect illustration of the role of kinetic time-symmetric aspects in stochastic transport.  We have followed the scheme of Astumian \cite{astu,ratastu} where the underlying microscopic reversibility leads to local detailed balance in the setting up of a Markov model. Asymmetry in the reactivities leads in the presence of excess ATP to forward motion.  The analysis of the corresponding dynamical ensemble allows to elucidate that frenetic contribution.  Using a mirror symmetry between trailing and leading heads we predict a fluctuation symmetry in the reflection antisymmetric part of the dynamical activity.  That fluctuation symmetry also determines the direction of the motor's motion.\\

\noindent {\bf Acknowledgment}:  We are very grateful to Karel Neto\v{c}n\'{y} for many discussions.  In particular WO'KdG thanks the Academy of Science and Karel Neto\v{c}n\'{y} for hospitality at the Institute of Physics in Prague.

\end{document}